\documentclass[aps,prb,superscriptaddress,preprint]{revtex4}
                                                                                
\usepackage{graphicx}% Include figure files
\usepackage{dcolumn}% Align table columns on decimal point
\usepackage{bm}% bold math
                                                                                
\begin{document}
\title{Reduction of the Three Dimensional Schr{\"o}dinger Equation
 for Multilayered Films}
\author{Charles Abbott}
\affiliation{U-46, Physics Department, University of Connecticut, Storrs, CT 06269}

\author{M. Rasamny}
\affiliation{Department of Computer and Information Sciences, Delaware State University,
             Dover, Delaware 19901}

\author{G. W. Fernando} 
\affiliation{U-46, Physics Department, University of Connecticut, Storrs, CT 06269}
\affiliation{Institute of Fundamental Studies, Hantana Road, Kandy, Sri Lanka}

\date{\today{}}
%\narrowtext
%\tighten
%\vskip 0.3in
%**********************
%*  Begin text        *
%**********************
\begin{abstract}

In this paper we present a method for reducing
 the three dimensional Schr{\"o}dinger equation to study
confined metallic  states, such as quantum well states, in a multilayer film geometry.
 While discussing some approximations that are employed
when dealing with the three dimensionality of the problem,
we derive a one dimensional equation suitable for studying
such states  using an envelope function approach.  Some applications
to the Cu/Co multilayer system with regard to
spin tunneling/rotations and angle resolved photoemission experiments
 are discussed.
\end{abstract}

%\pacs{}

\maketitle

\section{Introduction}
There has been much recent interest in magnetoelectronic
 devices due to their potential applications as 
miniaturized computer memory components and high speed 
analog devices~\cite{one,two}.  The ability to synthesize   
systems with artificial  structures has grown remarkably
over the past decade.  For example, it has been shown
 recently that molecular beam epitaxy techniques allow for the 
growth of independent ferromagnetic layers on a semiconducting
 substrate, in cases such as Fe on InAs(100)~\cite{two} 
and MnGa on GaAs(100)~\cite{four}. 
The search for new device materials and making optimum  use of
such new devices will greatly benefit from  an accurate understanding
 of the underlying quantum mechanical processes 
involved in electron transport as the dimensions of the device
 approach the wavelength of transmitting electrons.  
Recent experimental studies of
 spin dependent, hot electron transmission 
 such as those described in Ref.~\onlinecite{filipe} and
 resonant tunneling through two discrete states (Ref.~\onlinecite{vaart}),
 have raised a number of interesting issues related
to the ferromagnetic and insulating materials used, 
the nature of the electronic states that are involved in transmission
and enhancements in spin filtering effects. Apart from the first principles
 based attempts which can be quite tedious, most theoretical
 studies of these spin dependent effects have used free electron
band structures and other simplifications in the  metallic as well as in the
insulating  regions.
 Our work, though motivated by free electron
  approaches such as those introduced by Slonczewski~\cite{sol1}, is an
 attempt to bring out a more realistic lateral dependence of the
 electronic states under consideration.

  This paper is organized as follows. First,  starting from the
 three dimensional Schr{\"o}dinger equation, we proceed to derive
 an envelope function approach suitable for multilayered films.
 This procedure will go beyond the free electron methods that
 have been commonly used in the past,
 making use of more realistic wave functions, but
 avoiding a full pledged ab-initio calculation when studying such systems. 
 We introduce an approach
 which incorporates the two dimensional Bloch wave vector ${\bf k_{||}}$ 
 and show that  the associated parallel band structure characteristic 
 of the material being used, 
 plays a major role in perpendicular transmission. 

  Second, We will  study spin tunneling and rotation
 effects (to be defined later)
  in a multilayer system with two ferromagnetic layers
separated by a nonmagnetic metal such as Co/Cu/Co.
 We will also address some issues related to 
 angle resolved photoemission and inverse photoemission experiments 
 focused on  confined states in metallic multilayers. Although there have been  attempts
to explain such spectra using a Phase Analysis Model (PAM) with
some success~\cite{curti}, a better understanding of the multilayering,
tunneling, spin rotation effects etc. is necessary, for example,  to
go beyond simple situations and study more complicated heterostructures 
 of different sizes and shapes.
 For such systems, quantum mechanical calculations
 of spin-dependent transmission using a simple 
free-electron model have already been preformed~\cite{six,ten}.
However, there is a growing need for more realistic, yet simple
enough calculations in order to understand  new and small structures
that are being synthesized today.

\section{Model and Approximations }

       There have been some recent attempts  to 
evaluate the effects of the two dimensional (planar), metallic
 periodic structures on the confined states
 in various devices using simplified models~\cite{butler}.
   Some of these have focused on a Fourier space description 
of the one particle Schr{\"o}dinger wave function.
  We will be  examining some of the assumptions
 made in these Fourier as well as real space models about the confined
states, the effects of the planar regions on them and the importance
of spin asymmetry on the spin filtering process.

 Our model consists  of  multilayered slabs of different materials sandwiched
 together to form a device. For example, the device may contain
 several layers of Cu sandwiched between two ferromagnetic slabs of Co
 as in Co/Cu/Co. 
We will choose the $z$ direction to be perpendicular
 to these slabs and the $(x,y)$ to be parallel planes 
consisting of slabs. These can also
 be labeled longitudinal and transverse directions respectively.
   There are several simplifications that are usually made in attempts
 to calculate properties of such structures.
  When approximations are made to the
 wave function near an interface, it is a common practice to separate the
 transverse $(x,y)$ dependence and the longitudinal $(z)$ dependence (see
  Ref. \onlinecite{butler}).   
    Let us carefully consider the conditions on the one particle potential and
 the tunneling states that lead to such a  description of the problem
 at hand. An arbitrary eigenstate  here can be expressed in the (planar) Bloch form as

\begin{equation}
 \Psi_{\bf k_{||}}~ ({\bf r_{||}}, z) = {\sum_{\bf G_{||}}} C_{\bf G_{||}}^{\bf k_{||}} (z)~
 exp\{ i({\bf k_{||}} + {\bf G_{||}})
    \cdot {\bf r_{||}} \}  
\label{eq:wf}
\end{equation}
and the one particle potential $U(x,y,z)$ can  be expressed as

\begin{equation}
 U (x,y,z) = {\sum_{\bf G_{||}}} V_{\bf G_{||}} (z)
      ~exp(i{\bf G_{||}}\cdot {\bf r_{||}})  
\label{eq:pot}
\end{equation}
  to elucidate the periodic nature of the potential and 
  the Bloch-like form of the wave function
  in the planar (i.e. parallel) direction. Here ${\bf G_{||}}$ refers
 to a planar reciprocal lattice vector, while ${\bf r_{||}} = (x,y)$. 

  It is possible to analyze the effects of the planar states on the
 perpendicular behavior  in several different ways. First,  using
 the Fourier coefficients defined above, the complete one particle
 solution may be expressed as,

\begin{equation}
-\frac{\hbar^2}{2m}\frac{\partial^2}{\partial z^2}
C_{\bf G_{||}}^{\bf k_{||}}(z)
 + \sum_{{\bf G_{||}^{'}}\ne {\bf G_{||}}} V_{\bf G_{||} - G_{||}^{'}}(z)
 C_{\bf G_{||}^{'}}(z) = \{ E - \frac{\hbar^2}{2m}({\bf k_{||}} +
 {\bf G_{||}})^{2} - V_{0}(z)\} C_{\bf G_{||}}^{\bf k_{||}}(z)
\end{equation}  
Note that unless the (parallel and z direction) potential coefficients
$V_{\bf G_{||} - G_{||}^{'}}(z)$ are weak compared to the relevant
 energy scale of the problem, the above equation couples the Fourier
 coefficients of the wave function through the potential coefficients
 and hence does not necessarily yield exponentially
 decaying solutions in the perpendicular (z) direction
 for the wave function even if the  energy $E$ satisfies the 
 condition,

\begin{equation}
  E < \frac{\hbar^2}{2m}({\bf k_{||} +
 {\bf G_{||}})^{2}} + V_{0}(z)
\end{equation}
for all values of $z$.

  Consider the wave function as defined in Eq.\ (\ref{eq:wf}). 
 An assumption that is usually made~\cite{butler} when looking for 
 such solutions is the following "separability" condition: 
\begin{equation}
\Psi = \xi^{'}(x,y)\phi^{'}(z).
\label{eq:sep}
\end{equation}  
  The assumption of the wave function
 separating into  planar ($\xi^{'}$)
 and  perpendicular 
($\phi^{'}$) parts 
 is  equivalent to having
 $C_{\bf G_{||}}(z) = \phi^{'}(z) D_{\bf G_{||}}$, where $D_{\bf G_{||}}$
 Fourier coefficient has no $z$ dependence.
  These ideas can be expressed in terms of the one particle potential  $U(x,y,z)$
 and its Fourier
 expansion given by Eq.\ (\ref{eq:pot}).
 It is clear that if this potential satisfies the  (additivity) condition,
  $U(x,y,z) = U_{1}(x,y) + U_{2}(z)$ with no coupling between the planar $x,y$ and
 perpendicular $z$ dependencies, then the above separation of variables
in the wave function can be justified. In such situations, the Hamiltonian H
also becomes additive as $H(x,y,z) = H_1(x,y) + H_2(z)$ and a simple quantum well
equation,
\begin{equation}
\left[ -\frac{\hbar^2}{2m}\frac{\partial^2}{\partial z^2} + U_2(z)
\right]\phi^{'}(z) = E_C\phi^{'}(z)
\label{eq:1d}
\end{equation}
can be obtained for the $\phi^{'}(z)$.

    However, we argue that the above assumptions  are too restrictive when one is
 dealing with states that have $p$,  $d$ or $f$ character in local orbital angular 
 momentum. The coupling of the three directions in the wave functions have to be
 dealt with more rigor, since such atomic wave functions are unlikely to satisfy
 the separability condition expressed in Eq.\ (\ref{eq:sep}).  Significant corrections
 will be necessary if the separability condition is used as a starting point
 for better calculations. Hence we have sought a different starting point 
 for carrying out the reduction of the three dimensional Schr{\"o}dinger
 equation.  
 The  envelope function approach discussed below
 provides an ideal and formal platform for handling the situation at hand.

\section{Envelope Functions and the Full Problem}

 With  the advent of methods related to semiconductor quantum wells, 
 the simplistic theory surrounding
 them has been so successful that sometimes it is easy to undermine 
 its connections to the well understood regime of weak perturbations
 in bulk crystals. 
  However, since the potentials employed in quantum wells are strongly
 perturbed at various boundaries, some formal justification seems necessary
 in order to use simple quantum well equations for multilayer systems.
  In fact, some applications of quantum well based techniques  to semiconductor
 heterostructures have  been justified
 using an envelope function approach~\cite{burt}.

      In principle, the relevant many particle Hamiltonian
 in all the different regions of the heterojunction carries all the
 necessary (interaction) information. When combined with the proper boundary
 conditions, its appropriate eigenstates can be used to
 describe, for example, tunneling in such a device. However, this problem is highly
 nontrivial and various approximations are sought in  order to simplify it.
 First principles methods, such as those based on the density functional theory,
  can be utilized for this purpose but less complicated approaches
 that can reduce the computational burden are 
 quite attractive. The envelope function method introduced by Bastard~\cite{nine}
 is one such approach.
 Here the real space equations satisfied by the envelope functions were equivalent
 to the ${\bf k}\cdot{\bf p}$ method of Kane~\cite{kane} with the band edges 
 allowed to be functions of position.  For heterojunctions with 
 planar metallic regions, the applications of such ideas
 can be clarified and presented in a relatively straightforward way starting
 from the nonrelativistic three-dimensional Schr{\"o}dinger equation:

\begin{equation}
\left[ -\frac{\hbar}{2m} \nabla^2 + U_{\sigma}(x,y,z) 
\right]\Psi(x,y,z) = E \Psi(x,y,z).
\label{eq:s1}
\end{equation}                  
  As is commonly done for (nonrelativistic) itinerant systems, 
 the spin dependence in the Hamiltonian
 has been absorbed into a  spin dependent potential $U_{\sigma}$, and
 the two equations representing the up and down spins can be solved separately.
 Hence, from now on, we will drop the spin index $\sigma$ and
  focus on the reduction of a single, one particle Schr{\"o}dinger equation. 
In the parallel direction, the metallic as well as insulating regions are assumed
 to have perfect, two dimensional 
crystalline order, giving rise to extended electronic states with well defined parallel
 Bloch momenta $\hbar {\bf k}_{||}$.  We 
also assume perfect (parallel) lattice vector matches at  various interfaces.

      An important point to note here is that in
 these problems that involve heterojunctions,
 there are (at least) two relevant length scales;
 namely the interatomic length scale and the scale associated
 with the (confining) structures. The envelope function may vary on the latter
 length scale or on some combination of the two, which is to be determined later.
 Based on this argument, one can expect a nontrivial envelope function, when it exists,
 to modify the rapidly varying atomic wave functions. We regard this as our
 starting point and  express the full problem
 (ignoring the spin dependence) as:

\begin{equation}
\left(-\frac{\hbar^2}{2m} \frac{\partial^2}{\partial z^2} +
\left[-\frac{\hbar^2}{2m}\left(\frac{\partial^2}{\partial x^2} +
\frac{\partial^2}{\partial y^2} \right) + U(x,y,z) 
  \right]\right) \xi(x,y,z)\phi(z) =
 E\xi(x,y,z)\phi(z).
\label{eq:s2}
\end{equation}

    The function $\xi(x,y,z)$ can be thought of as a wave function 
with rapid variations on the atomic scale that has the
 two dimensional Bloch character, while $\phi(z)$ is an envelope function 
 as described above.  An important point to
  note here is that we do not make an assumption on separability as in Eq. (\ref{eq:sep}).
The existence of a nontrivial envelope function, as identified by the above equation,
will be used as a prerqusite for the existence of quantum well or other confined states.
 Our search is for envelope functions, as defined above, that are likely
 to arise due to the confining structures. 
 One can question the validity of such an
 expression, and  similar forms have been suggested~\cite{burt} such as
 an expansion using products of Bloch states and envelope functions.
  Here, we use the above form
  and associate an eigenvalue $E_{||}$ with the function
 $\xi(x,y,z)$ through the following eigenvalue problem.

\begin{equation}
\left(-\frac{\hbar^2}{2m} \left[\frac{\partial^2}{\partial z^2} +
\left(\frac{\partial^2}{\partial x^2} +
\frac{\partial^2}{\partial y^2} \right)\right] + U(x,y,z)   \right) \xi(x,y,z) =
 E_{||}\xi(x,y,z)
\label{eq:s3}
\end{equation}

   We note that this is similar to how  two dimensional band structure
 is calculated for thin films. Although for simplicity, we do not address
 interface roughness and other similar issues in this paper, the potential $U(x,y,z)$
 in Eq. (\ref{eq:s2}) can be modified to include such
 effects with appropriate changes in the
 boundary conditions (and a z dependent potential in Eq. (\ref{eq:pz})).
   With the above definition of $E_{||}$ (subband energy),  the
 full Schr{\"o}dinger equation can be used to obtain a differential
  equation for the envelope function $\phi (z)$ in the following manner.

\begin{equation}
-\frac{\hbar^2}{2m}\left[\xi(x,y,z) \frac{\partial^2}{\partial z^2} \phi(z) +
 2\frac{\partial}{\partial z}
\phi(z) \frac{\partial}{\partial z}\xi(x,y,z) \right] =
 (E-E_{||})\xi(x,y,z)\phi(z)
\label{eq:wfz}
\end{equation}

     The second  term on the left hand side
 of the above equation contains a product of first order derivatives
 and describes  some coupling of the lateral (i.e. planar)  and
 longitudinal (i.e. perpendicular) coordinates, in addition to
 any coupling that is already contained in $E_{||}$.
  This coupling term can be simplified using averages of
 $\xi(x,y,z)$ over the planar $x,y$ coordinates. In general,
 this will result in a $z$ dependent term and a second order differential
 equation for $\phi(z)$ as, 
\begin{equation}
-\frac{\hbar^2}{2m}\left[\frac{\partial^2}{\partial z^2} \phi(z) +
  P_{||}(z)\frac{\partial}{\partial z}
\phi(z)  \right] =
 (E-E_{||})\phi(z)
\label{eq:pz}
\end{equation}
with
\begin{equation}
 P_{||}(z) = (2
\int\int dx dy ~~\xi^{*}
 \frac{\partial \xi}{\partial z}) ~/~  
(\int\int dx dy ~~\xi^{*}\xi).
\label{eq:pz1}
\end{equation}
The double integral over the planar coordinates $x,y$ has to be carried
out over a suitable 2 dimensional unit cell.
 This is a mathematically  rigorous result, based solely on the
 assumptions stated previously,
 and the theory at this level cannot and should not
distinguish between metals, semiconductors or insulators.

  Now we can look for possible simplifications to Eq. (\ref{eq:pz})
 by monitoring the properties of  $P_{||}(z)$. Note that if the
 Bloch function $\xi (x,y,z)$ has an oscillatory $z$
 dependence, then $P_{||}(z)$ will be purely imaginary.
 However, the imaginary part of $P_{||}(z)$ ($Im~P(z)$) is directly related
 to the $(x,y)$ averaged flux, $J(z)$ along the $z$ direction. Hence,  we can make use of
 (steady state) flux
 conservation which leads to ${\partial J(z) }/{\partial z}  = 0$; i.e.
 the conservation
 of flux implies that $Im~(P_{||})$ has to be independent of $z$. When $P_{||}(z)$
 has a non negligible real part, we cannot make use of the above argument,
 and the type of confined states that we are searching for
 (through Eq. (\ref{eq:mqw1}) ), will not exist.
 However, note that even for this situation, we have achieved the
 reduction of the 3 dimensional Schr{\"o}dinger equation to a one
 dimensional one.
 Now setting $P_{||}(z)=Q_{||}$ (z-independent), the substitution
\begin{equation}
\phi(z) = exp(-zQ_{||}/2)\zeta(z)
\end{equation}
 can be used to eliminate the first order derivative
leading to a familiar equation,
similar to Eq. \ (\ref{eq:1d}). The function $\zeta (z)$ appearing here
 has a different interpretation, as a part of {\sl an envelope function}, and
 the boundary conditions in the quantum well problem should be applied to
 the envelope function $\phi(z)$ or $\zeta(z)$.  Note that that  if
 $\zeta(z)$ and its derivative are continuous across various boundaries along
 the $z$ direction, then similar properties can be established for $\phi(z)$.

\begin{equation}
-\frac{\hbar^2}{2m}\left[\frac{\partial^2}{\partial z^2} \right]\zeta(z)=
 \left[E-E_{||}({\bf k_{||}}) -\frac{\hbar^2Q_{||}^2}{8m}\right]\zeta(z) = E_C\zeta(z)
\label{eq:mqw1}
\end{equation}

  The above equation illustrates several important points that are often
overlooked or misinterpreted in simplistic quantum well  and free electron approaches.
As observed in Ref. (\onlinecite{butler}), a possible reduction in tunneling
rates can be expected due to the lateral variation of the wave function. Although
 our method is somewhat different,
 note that Eq. (4) of Ref. (\onlinecite{butler}) has a $(x,y)$ averaged term similar to
 $Q$ in our discussion, which affects, for example, decay rates associated with 
 $s,~p$ functions  differently. 
 The subband structure, i.e. $E_{||} ({\bf k_{||}})$,
in the multilayers affect the confined eigenstates and eigenvalues.
The envelope functions are also affected by the averages of the parallel Bloch
functions through $Q_{||}$.
  For each `confined energy level' with energy $E_{C}$, there exists a
continuous subband of states that share the same {\sl perpendicular} wave function $\phi(z)$
but differ in {\sl parallel} Bloch momentum ${\bf k_{||}}$, $E_{||}$
 and $\xi (x,y,z)$.
Note that $E_{||}$ can  depend on the thickness of a given multilayer
film and carries information not only about the planar
structures, but also about the longitudinal coupling, following
our definition through Eq. ({\ref{eq:s3}).
 Finally, the energy $E$ of a given
electron in such a quantum mechanical state depends on all of the above.

\section{Applications - Confined States in Metallic Multilayers}

 We are now ready to apply our model to test devices.
 The first device consists of several layers of (nonmagnetic) Cu, sandwiched between two
 slabs of (ferromagnetic) Co  (i.e. Co/Cu/Co) which will be used to discuss
spin transmission/rotation effects. 
 For an incident electron of energy $E_{total}$,  
 we obtain the following electronic perpendicular momenta,

\begin{eqnarray}
\hbar k_{\uparrow z} = \sqrt{2m(E_{total}-E_{||}) - \hbar^2 Q_{||}^2/4} \\
\hbar k_{\downarrow z}^{\Delta} = \sqrt{2m(E_{total}-E_{||}-\Delta ) - \hbar^2 Q_{||}^2/4}
\label{eq:mqw0}
\end{eqnarray}
where $k_{\uparrow }$ ($k_{\downarrow}^{\Delta}$) is the perpendicular wave vector
 for majority (minority) electrons in the 
 ferromagnetic, metallic regions where
the lateral effects have been taken in to account using the ideas
developed in the previous sections. Here $\Delta$ is the spin splitting
 in the two dimensional bands assumed to be {\bf k$_{||}$} independent.   

In this device, we  can 
 rotate the magnetization of the right (R) ferromagnet by an angle $\theta$ with 
respect to the magnetization of the left (L) ferromagnet. The spin rotation
is introduced at the boundary $z=0$ inside the spacer (Cu) layer for simplicity.
For this device, solutions 
 to Eq. (\ref{eq:mqw1})  in various spatial
regions (L and R, subdivided into $a,c,d,e$) take the form,

\begin{eqnarray}
  \left[ 
     \begin{array}{c}
       \zeta^{L}_{a\uparrow} \\ \zeta^{L}_{a\downarrow}
     \end{array} 
   \right] 
   &  = & 
   \left[ 
      \begin{array}{c} 
          Ie^{i k_{\uparrow}^a z} +  R_{\uparrow\uparrow}e^{-i k_{\uparrow}^a z} \\ 
          R_{\uparrow\downarrow}e^{-i k_{\downarrow}^{\Delta_a }z} 
      \end{array} 
   \right] \nonumber \\
%=============
   \left[ 
      \begin{array}{c} 
         \zeta^{L}_{c\uparrow} \\ \zeta^{L}_{c\downarrow}
      \end{array}
   \right]
   & = &
   \left[ 
      \begin{array}{c} 
         C_{\uparrow +}e^{i k_\uparrow^c z} + C_{\uparrow -}e^{-i k_\uparrow^c z} \\
         C_{\downarrow +}e^{i k_{\downarrow}^c z} + C_{\downarrow -}e^{-i k_{\downarrow}^c z}
      \end{array} 
   \right] \nonumber \\
   \left[ 
      \begin{array}{c} 
         \zeta^{R}_{d\uparrow} \\ \zeta^{R}_{d\downarrow}
      \end{array}
   \right]
   &  = & 
   \left[
      \begin{array}{c} 
         D_{\uparrow +}e^{i k_\uparrow^d z} + D_{\uparrow -}e^{-i k_\uparrow^d z} \\
         D_{\downarrow +}e^{i k_{\downarrow}^d z} + D_{\downarrow -}e^{-i k_{\downarrow}^d z}
      \end{array} 
   \right] \nonumber \\
%=============
   \left[ 
      \begin{array}{c} 
         \zeta^{R}_{e\uparrow} \\ \zeta^{R}_{e\downarrow}
      \end{array}
   \right]
   & = &
   \left[ 
      \begin{array}{c} 
         T_{\uparrow\uparrow}e^{i k_\uparrow^e z}   \\
         T_{\uparrow\downarrow}e^{i k_{\downarrow}^{\Delta_e }z} 
      \end{array} 
   \right].
\label{eq:mqw2}
\end{eqnarray}
Finally, we impose the boundary conditions at $z \pm z_2$ (where $2z_2$ is the
thickness of  the center slab) and at $z =0$ for spin rotations.

\begin{eqnarray}
  \left[ 
     \begin{array}{c} 
        \zeta^{L}_{a\uparrow}(-z_2) \\
        \zeta^{L}_{a\downarrow}(-z_2) 
     \end{array} 
  \right]
  & = & 
  \left[ 
     \begin{array}{c} 
        \zeta^{L}_{c\uparrow}(-z_2) \\
        \zeta^{L}_{c\downarrow}(-z_2) 
     \end{array} 
  \right] \nonumber \\
%===============
  \left[ 
     \begin{array}{c} 
        \zeta^{R}_{d\uparrow}(0) \\
        \zeta^{R}_{d\downarrow}(0) 
     \end{array} 
  \right]
  & = & 
  S(\theta)
  \left[ 
     \begin{array}{c} 
        \zeta^{L}_{c\uparrow}(0) \\
        \zeta^{L}_{c\downarrow}(0) 
     \end{array} 
  \right] \nonumber \\
%================
  \left[ 
     \begin{array}{c} 
        \zeta^{R}_{d\uparrow}(z_2) \\
        \zeta^{R}_{d\downarrow}(z_2) 
     \end{array} 
  \right]
  & = & 
  \left[ 
    \begin{array}{c} 
       \zeta^{R}_{e\uparrow}(z_2) \\
       \zeta^{R}_{e\downarrow}(z_2) 
    \end{array} 
  \right]
\label{eq:mqw3}
\end{eqnarray}
and identically for the first derivative.  
The matrix $ S(\theta)=\left( \begin{array}{cc}cos(\theta/2)&sin(\theta/2)\\
-sin(\theta/2)&cos(\theta/2)\end{array}\right)$ 
is the spinor transformation, 
where $\theta\ne 0$ is tied to the spin rotation effects as discussed below.  
We can then fully determine the transmission coefficients utilizing
 numerical techniques.

\subsection{ Spin Transmission and Rotations}

 An interesting application of the above ideas has to do with ballistic 
 transport and spin rotation effects.
  When polarized electrons are transmitted from one region to
 another with a different polarization axis, they experience
 a spin-torque and a transfer of angular momentum to the new
 medium~\cite{sol2}.  These ideas have now been demonstrated
 experimentally, for example, through the phenomenon of giant
 magnetoresistance where large current densities
 flowing perpendicular to the films have been observed in
 reversals of magnetization~\cite{myers}, and spin precession~\cite{web}.
 This field is an emerging one related to `magnetoelectronics' and 
 many new experiments are expected to be conducted on spin
 transmission/rotation effects in magnetic multilayer systems.

 We do not wish to focus on the mechanisms of spin transfer but 
 simply use spin rotation angle as an input to our calculations
 and obtain the corresponding transmission coefficients in 
 a device consisting of two magnetic films separated by a nonmagnetic one
 (i.e. Co/Cu/Co). Spin rotation effect is introduced, by hand, at the center
 of the nonmagnetic film (Cu) for simplicity as has been done previously~\cite{sol1}.
 However, unlike in the previous studies,
 the underlying band structure and lateral effects have been taken into account
 by using the theory of confined
 states developed here. These band dispersions
 play a crucial role in determining the spin dependent transport
 properties. For example, 
 when such confined  states in a ferromagnetic film are located in a gap
 of minority spin states,
 the system can act as an almost perfect spin filter .
  The k-dependent
  transmission coefficients, $
         T_{\uparrow\uparrow}$ and
        $ T_{\uparrow\downarrow}, $
  that were introduced  in the previous section
  can be used through a  Landauer type formula~\cite{lan},
\begin{equation}
 G = {e^2\over h}\sum_{\bf k_{||}}~T({\bf k_{||}})
\label{eq:land}
\end{equation}
 to obtain the
  conductivity in the quantum well problem under discussion.
  Here we calculate transmission coefficients along $\bar\Gamma\bar X$
 using selected subbands, as defined in Eq. (\ref{eq:band}),
\begin{equation}
E_{||}({\bf k_{||}})=E_{||}^0~+~W\{ 1 - cos(k_{x}a) \}.
\label{eq:band}
\end{equation}
 in the transmission device (Co/Cu/Co) for the following
 set of parameters (Fig. \ref{fig:mqw1}) for illustrational
 purposes: W(Co)= 0.07, W(Cu)= 0.06, $\Delta$ = 0.19 (all in Rydbergs),
 $z_2$= 13.6, a= 6.8 (all in Bohrs).

\begin{figure}
\includegraphics*[width=20pc]{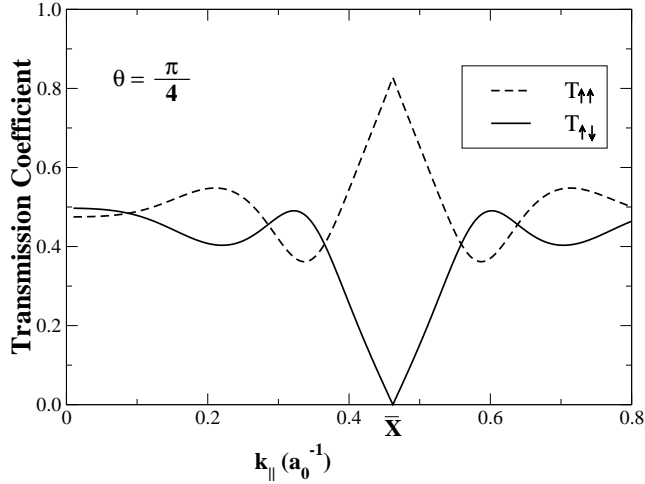}
\caption{Transmission coefficients $T_{\uparrow\uparrow}, T_{\uparrow\downarrow}$
         as a function of ${\bf k_{||}}$ along $\bar\Gamma\bar X$ in the second
         device (Co/Cu/Co) for a relative spin orientation $\theta=\frac{\pi}{4}$. 
         At the zone boundary ${\bar X}$, the device filters out the
   `minority' spins.  
         Away from the zone boundary both
        `majority' and `minority' spins are transmitted.}
\label{fig:mqw1}
\end{figure}

 At a given (total) energy of the incoming electron, and a given
 relative spin orientation of the Co films, transmission coefficients
 $T_{\uparrow\uparrow}, T_{\uparrow\downarrow}$ have been obtained
 from Eqs. (\ref{eq:mqw2}) and (\ref{eq:mqw3}) as a function of
 the parallel Bloch momentum, ${\bf k_{||}}$.
 From this figure, it is easy to  see the effects of
 different types of Bloch states on the tunneling. Near the
 zone boundary ${\bar X}$, for the given relative spin orientation $\theta$,
 we see the device filtering out  `minority' spins, while at other
 k-points along $\bar\Gamma\bar X$, a mixture of both `majority' and
 `minority' spins are transmitted. This is a direct result of the
 upward dispersion of the selected subbands along $\bar\Gamma\bar X$ and
 the spin splitting in Co, pushing the minority band closer to
 the given (total) energy of the electron.

\subsection{Energy Spectrum and  Photoemission}

 Angle Resolved Photoemission (ARP)
 techniques have become a useful tool for studying
electronic states and their (band) dispersions in structures with
two dimensional (planar) Bloch symmetry. In ARP experiments
 photons are used to eject electrons from occupied states
while in inverse ARP, photons are ejected, when the above photoemission
 process is reversed. 
 Photoemission is a many body phenomenon and
an evaluation of the spectral function and appropriate self-energies
yield the full photoemission spectrum and peak widths. However,
here we follow a simpler approach and focus on one particle energy
 states and the changes introduced by the multilayering. In this approach,
  the energy $E$ that is measured for the photoemitted electron
can be associated with the full Schr{\"o}dinger equation (\ref{eq:s1}).
Existence of an envelope function as defined in Eq. (\ref{eq:s2}) and
satisfying the imposed boundary conditions are necessary for confined states. 
	 The energy $E_{||}$ as defined through Eq. (\ref{eq:s3}) has to do with
the ordinary subband states due to the periodic potential in the planar directions.
If the multilayering does not play any role for a
 given energy $E$ at a given ${\bf k_{||}}$, then
$E_{||}$ and $E$ have to be identical when these solutions exist in Eqs. (\ref{eq:s1})
and (\ref{eq:s3}). In such cases, $\phi(z)$ turns out to be a simple multiplicative
constant and Eq. (\ref{eq:wfz}) is consistent with this scenario. However, when
confined states exist, they may alter the usual
 dispersions $E_{||}$ that are observed in their absence.
   According to the theory developed above, confined states can be
identified as states for which solutions to the differential
equation (\ref{eq:mqw1}) exist for a given energy $E$ of the electron.
 It is also important to realize that, for a given energy $E$,
 these states may not exist for all values of ${\bf k_{||}}$ along a
given direction of the 2 dimensional Brillouin zone.

 In general, the dispersion of the total energy of the electron
$E$, i.e. the energy of the bound electron as a function of
${\bf k_{||}}$, are found 
whenever  confined states exist with appropriate $ \zeta (z)$ 
 as solutions to Eq. (\ref{eq:mqw1}).  In the present context,
 with boundary conditions appropriate to stationary states
 (determined by the film geometry), we obtain a set of discrete
 energies ($E_n$) and states
 for the one dimensional quantum well or barrier problem where
  
\begin{equation}
E = E_n
 + E_{||}({\bf k_{||}}) + \frac{\hbar^2Q_{||}^2}{8m}.
\label{eq:mqw5}
\end{equation}
 Hence, the observed energies $E$ in an ARP experiment for confined states will
 depend on the existence of these discrete $E_n$s and the corresponding
 $\zeta$ functions satisfying the relevant boundary conditions.
 As in simple quantum well problems, solutions of odd and even parity
 may sometimes be associated with $E_n$ and these will depend on the
 film geometry, interfaces and growth conditions. We plan to address these as
 well as effects due to film thickness (observed in Refs. \onlinecite{curti,wu} )
 in a future paper.

\section{Conclusions}

 Using an envelope function approach,
we have reduced the three dimensional Schr{\"o}dinger equation
to a one dimensional one and  analyzed the lateral effects of metallic multilayer films
through a hypothetical subband structure that varies with the two dimensional Bloch
vector.  Ballistic transmission
in a magnetoelectronic device consisting of
 Co/Cu/Co has been examined and shows  a clear
dependence on the  Bloch vector ${\bf k_{||}}$.
 Energy spectrum of confined  multilayer states that may be observed in
angle resolved photoemission studies are discussed. 
 Also, this method of reduction
of the Schr{\"o}dinger equation with appropriate modifications may
be used in other confined geometries such as nanoparticles and wires.

\section{Acknowledgements}
The authors wish to thank Dr. B. Sinkovic for useful discussions.
  This work has been supported by the National 
Science Foundation REU program at the University of Connecticut.

%*************************
%*  End text             *
%*************************


\begin{thebibliography}{10}

\bibitem{one}
{\it Physics Today} {\bf 48} (1995) pp.24-63.
\bibitem{two}
Y.B. Xu, D.J. Freeland, E.T.M. Kernohan, W.Y. Lee, M. Tselepi,
 C.M. Guertler, C.A.F. Vaz, J.A.C. Bland, S.N. Holmes, N.K. Patel and D.A. Ritchie, 
Jour. Appl. Phys. {\bf 85}, 5369 (1999).

\bibitem{four}
W. van Roy, H. Akinaga, S. Miyanishi and K. Tanaka, Appl. Phys. Lett. {\bf 69}, 711 (1996).

\bibitem{filipe} A. Filipe, H. -J. Drouhin, G. Lampel, Y. Lassailly,
  J. Nagle, J. Peretti, V. I. Safarov, and A. Schuhl, Phys. Rev. Lett.,
 80, 2425, (1998); D. J. Monsma, J. C. Lodder, T. J. A. Popma, B. Dieny,
 {\sl ibid}, 74, 5260 (1995).

\bibitem{vaart} N. C. van der Vaart, S. F. Godijn, Y. V. Nazarov,
C. J. P. M. Harmans, J. E. Mooij, L. W. Molenkamp, C. T. Foxon,
Phys. Rev. Lett., 74, 4702 (1995).

\bibitem{sol1}
J.C. Slonczewski, Phys. Rev. B {\bf 39}, 6995 (1989).

\bibitem{six}
Z. Zheng,Y. Qi, D.Y. Xing and J. Dong, Phys. Rev. B {\bf 59}, 14505 (1999).

\bibitem{curti}  F. G. Curti, A. Danese, and R. A. Bartynski, Phys. Rev. Lett.,
                {\bf 80}, 2213 (1998). 


\bibitem{five}
J.E. Ortega and F.J. Himpsel, Phys. Rev. Lett. {\bf 69}, 844 (1992).


\bibitem{seven} F. T. Vasko and A. V. Kuznetsov in Electronic States and
Optical Transitions in Semiconductor Heterostructures, Springer (1998).

\bibitem{eight}
J.P. Sun, G.I. Haddad, P. Mazumder and J.N. Schulman, Proc. of the IEEE, {\bf 86}, (1998).

\bibitem{nine} G. Bastard, Phys. Rev. B{\bf 24}, 5693 (1981);
  G. Bastard, J. A. Brum and R. Ferreira, Solid State Physics, Vol. 44,
  Academic Press, New York, p229 (1991). 

\bibitem{ten} T. Hayashi, M. Tanaka, and A. Asamitsu, Jour. Appl. Phys.
 {\bf 87}, 4673 (2000).

\bibitem{burt} M. G. Burt, J. Phys. Condens. Matt. Phys. R53 (1999);
              M. G. Burt, {\sl ibid}, {\bf 4}, 6651 (1991).

\bibitem{kane} E. O. Kane, Semiconductors and Semimetals, Vol. 1, p 75,
 Academic Press, New York (1975).

\bibitem{butler} W. H. Butler, X.-G. Zhang, and T. C. Schulthess and J. M. MacLaren, Phys.
                 Rev. B {\bf 63}, 092402 (2001).


\bibitem{sol2} J. C. Slonczewski, J. Magn. Magn. Mater., 159, L1 (1996); 195,
 L261 (1999); L. Berger, Phys. Rev. B 54, 9353 (1996).

\bibitem{myers} E. B. Myers, D. C. Ralph, J. A. Katine, R. N. Louie and
               R. A. Buhrman, Science 285, 867 (1999).

\bibitem{web} W. Weber, S. Rieseen, and H. C. Siegmann, Science 291, 1015 (2001).

\bibitem{lan} R. Landauer, Philos. Mag. 21, 863 (1970); M. B{\"u}ttiker, IBM Res. Dev.
              {\bf 32}, 317 (1988). 

\bibitem{wu} Y. Z. Wu, C. Y. Won, E. Rotenberg, H. W. Zhao, F. Toyoma,
                N. V. Smith and Z. Q. Qiu, Phys. Rev. B {\bf 66},245418 (2002).
\end{thebibliography}
\end{document}